\documentclass{kluwer}
\usepackage{epsfig}
\begin{document}
\begin{article}

\newcommand\apj[6] { #1: #2, '#3', {\it Ap.~J. \/} {\bf#4}, #5--#6.}
\newcommand\mnras[6] { #1: #2, '#3', {\it M.~N.~R.~A.~S. \/} {\bf#4},
  #5--#6.}
\newcommand\aap[6] { #1: #2, '#3', {\it A.~A. \/} {\bf#4}, #5--#6.}
\newcommand\icarus[6] { #1: #2, '#3', {\it Icarus \/} {\bf#4}, #5--#6.}

\newcommand\wig[1]{\mathrel{\hbox{\hbox to 0pt{%
          \lower.5ex\hbox{$\sim$}\hss}\raise.4ex\hbox{$#1$}}}}

\begin{opening}

\runningtitle{Disks, Extrasolar Planets and Migration}
\runningauthor{Terquem, Papaloizou \& Nelson}
 
\title{{ Disks, Extrasolar Planets and Migration}}
   
\author{{ Caroline \surname{Terquem}}}
\institute{Lick Observatory, University of California, Santa Cruz,
  CA~95064, USA}

\author{{ John C. B. \surname{Papaloizou}}}
\author{{ Richard P. \surname{Nelson}}}
\institute{Astronomy Unit, Queen Mary \& Westfield College, Mile End Road,
  London~E14NS, UK}
 
\author{\small{\it To appear in {\it 'From Dust to Terrestrial Planets'},
    eds W. Benz, R. Kallenbach, G.  Lugmair, \& F. Podosek, ISSI Space
    Sciences Series No. 9, reprinted in Space Science Reviews (January 1,
    2000) }} 
\institute{}

\begin{abstract}

We review results about protoplanetary disk models, protoplanet migration
and formation of giant planets  with migrating cores.

We first model the protoplanetary nebula as an $\alpha$--accretion disk and
present steady state calculations for different values of $\alpha$ and gas
accretion rate through the disk.  We then review the current theories of
protoplanet migration in the context of these models, focusing on the
gaseous disk--protoplanet tidal interaction.  According to these theories,
the migration timescale may be shorter than the planetary formation
timescale.  Therefore we investigate planet formation in the context of a
migrating core, consi\-dering both the growth of the core and the build--up
of the envelope in the course of the migration.

\end{abstract}

 
\end{opening}

\section{Disk Models}
\label{sec:disks}

Most theoretical protostellar disk models have relied on the
$\alpha$--para\-me\-tri\-za\-tion of the anomalous turbulent viscosity
proposed by Shakura~\& Sunyaev (1973).  In this context, the kinematic
turbulent viscosity $\nu$ is written $\nu = \alpha c_s H$, where $c_s$ is
the sound speed, $H$ is the disk scale height and $\alpha$ is a dimensional
parameter ($\alpha < 1$).  So far, only MHD instabilities (Balbus \& Hawley
1991) have been shown to be able to produce and sustain turbulence in
accretion disks.  However, because they develop only in adequately ionized
fluid, the parameter $\alpha$ is probably not constant through protostellar
disks, and it may even be that only parts of these disks can be described
using this $\alpha$ prescription (Balbus \& Hawley 1998 and this volume).
However, for the purpose of considering planet formation, as we are
interested in here, we will use such models.  Since they have already been
described in previous papers, we will not present them in detail here
again.  Instead we refer the reader to Papaloizou \& Terquem (1999) for
references and a detailed description.  Below we shall make use of the
results of some of the numerical calculations they
performed. \\

{\it Vertical Structure: \/} We consider thin disks which are in Keplerian
rotation around a star of mass $M_{\ast}=1$~M$_{\odot}$.  The opacity in
the disk, taken from Bell \& Lin (1994), has contributions from dust
grains, molecules, atoms and ions.  We are free to choose two parameters to
construct $\alpha$--disk models.  We take $\alpha$ and the local mass flow
rate through the disk which is defined as $\dot{M}_{st} \equiv 3 \pi
\langle {\nu} \rangle \Sigma$, where $\langle{\nu} \rangle$ is the
vertically averaged kinematic viscosity and $\Sigma$ is the surface mass
density of gas.  If the disk were in a steady state, $\dot{M}_{st}$ would
not vary with radius and would be the constant accretion rate through the
disk.  We note that $\dot{M}_{st}$ is related to $\alpha$ through $\nu$,
but it also depends on the distribution of mass in the disk.  At a given
radius $r$ and for given values of $\dot{M}_{st}$ and $\alpha$, we solve
the differential equations describing the disk vertical structure
(equations of vertical hydrostatic equilibrium, energy conservation and
radiative transport) with appropriate boundary conditions to find the
dependence of the temperature, pressure, mass density and radiative flux on
the vertical coordinate.  An important point to note is that as well as
finding the disk structure, we also determine $\Sigma$ for a given
$\dot{M}_{st}$, so that a relation between $\langle \nu \rangle$ and
$\Sigma$ is derived.  In order to investigate the solutions of the
diffusion equation which governs the disk evolution (see below) and for
other purposes, we found it convenient to derive analytic piece--wise power
law fits to this relation.  Details of these fits, which can be used for a
very wide range of disk parameters, are given in Papaloizou~\& Terquem
(1999). \\

{\it Time Dependent Evolution and Quasi--Steady States: \/} In general, an
accretion disk is not in a steady state but evolves diffusely according to
the following equation (see Lynden--Bell~\& Pringle 1974, Papaloizou~\& Lin
1995 and references therein):

\begin{equation}
{\partial \Sigma \over \partial t} = {3 \over r}{\partial \over
\partial r} \left[ r^{1/2} {\partial \over \partial r} \left( \Sigma
\langle \nu \rangle r^{1/2}\right) \right]
\label{aa4} .
\end{equation}         

\noindent  From this we see that the characteristic diffusion timescale 
at radius $r$ is $t_{\nu} \sim \left( r / H \right)^2 \Omega^{-1} / (3
\alpha )$, where $H$ is the disk semithickness.  For disks with
approximately constant aspect ratio $H/r,$ as applies to the models
considered here, $t_{\nu}$ scales as the local orbital period.  One thus
expects that the inner regions relax relatively quickly to a quasi--steady
state which adjusts its accretion rate according to the more slowly
evolving outer parts (see Lynden--Bell~\& Pringle 1974 and Lin~\&
Papaloizou 1985).  For estimated sizes of protostellar disks of about 50~AU
(Beckwith~\& Sargent 1996), the evolutionary timescale associated with the
outer parts is about 30 times longer than that associated with the inner
parts with $r < 5$~AU, which we consider here in the context of planetary
formation.  Thus these inner regions are expected to be in a quasi--steady
state through most of the disk lifetime. \\

\begin{figure}
\centerline{\epsfig{file=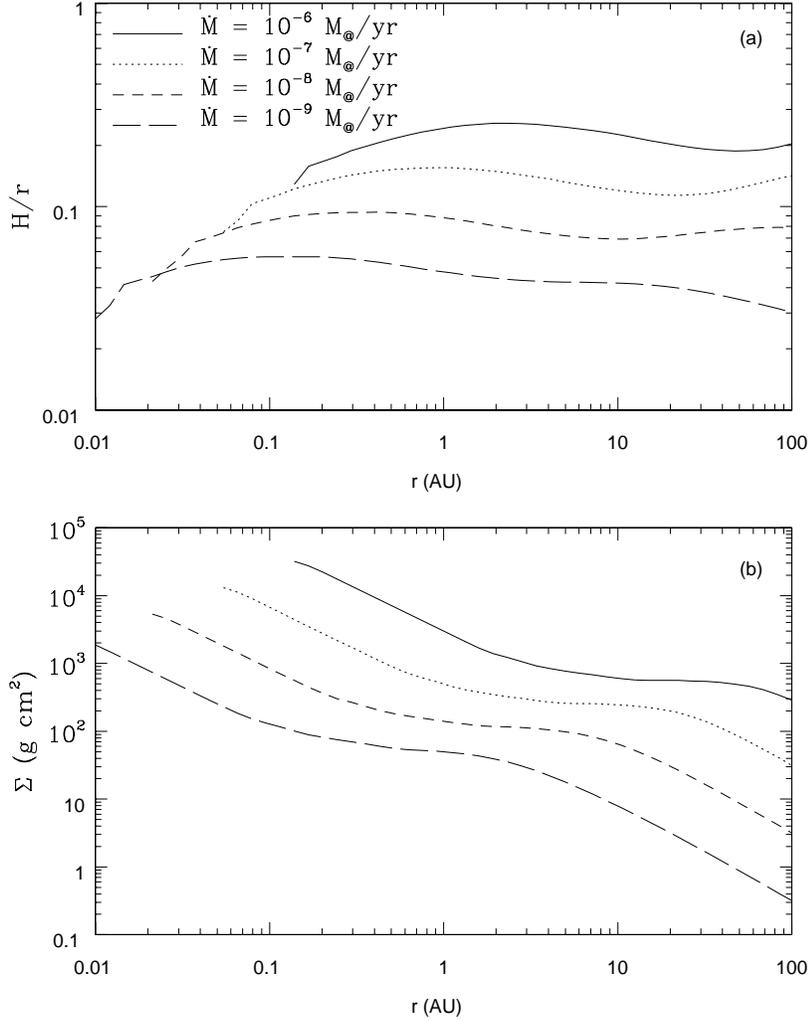,width=11.cm}}
\caption[]{Shown is $H/r$ ({\it plot~a}) and $\Sigma$ in 
  g~cm$^{-2}$ ({\it plot~b}) vs. $r/{\rm AU}$ using a logarithmic scale,
  for $\dot{M}_{st}$~(in units M$_{\odot}$~yr$^{-1}$)~$=10^{-6}$ ({\it
    solid line}), $10^{-7}$ ({\it dotted line}), $10^{-8}$ ({\it
    short--dashed line}) and $10^{-9}$ ({\it long--dashed line}) and for
  $\alpha=10^{-2}$.}
\label{fig1}
\end{figure}

\begin{figure}
\centerline{\epsfig{file=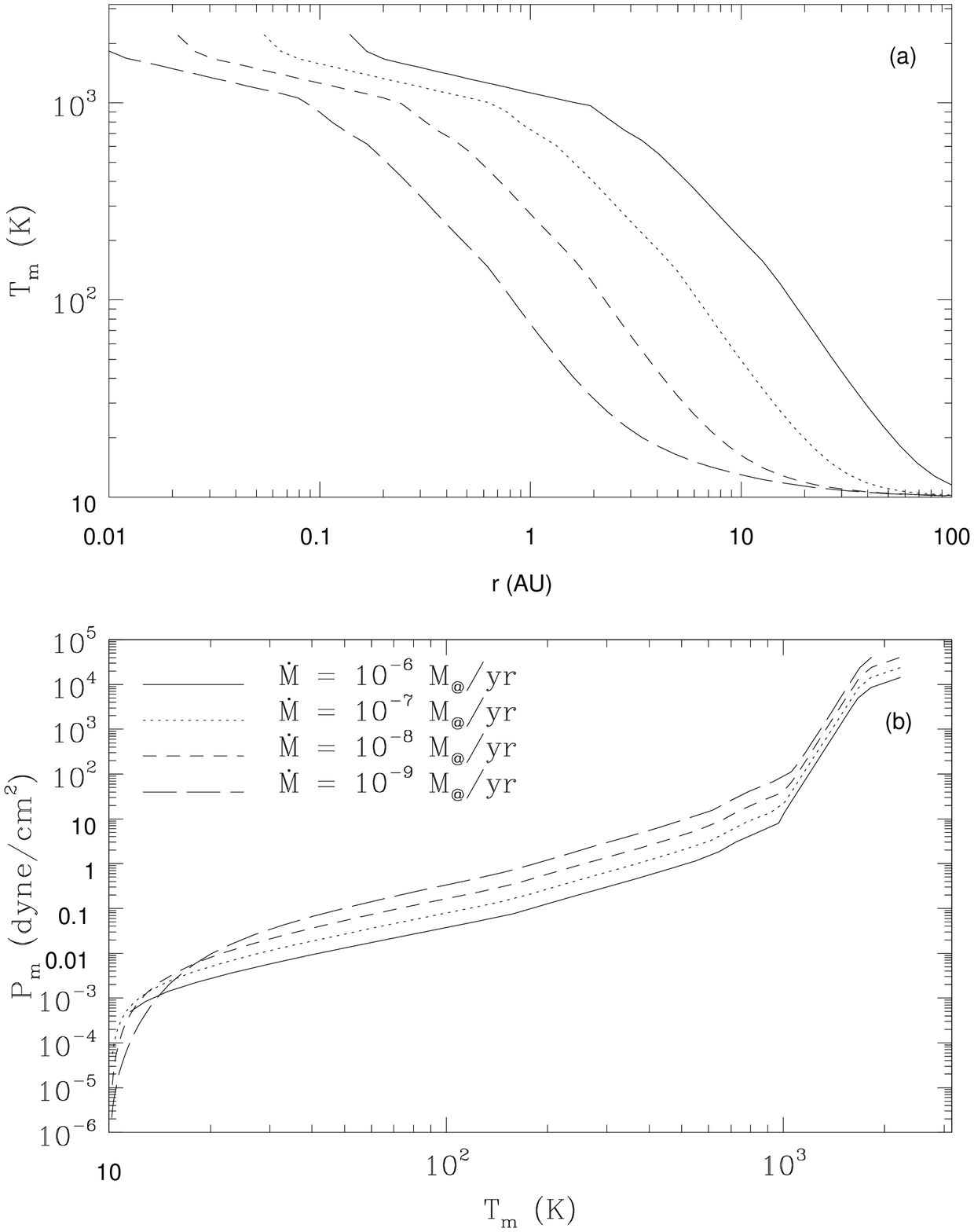,width=11.cm}}
\caption[]{Shown is $T_m$ in K ({\it plot~a}) vs. $r/{\rm AU}$
  and $T_m$ in K vs. $P_m$ in dyne~cm$^{-2}$ ({\it plot~b}) using a
  logarithmic scale, for the same parameters as in Fig.~\ref{fig1}.}
\label{fig2}
\end{figure}

{\it Steady State Models: \/} In the models we present here, we assume that
the disk is immersed in a medium with background temperature $T_b=10$~K.
In Figures~\ref{fig1}a--b we plot $H/r$ (where $H$ is defined as the
vertical height at approximately unit optical depth above which the
atmosphere is isothermal) and $\Sigma$ versus $r$ for $\dot{M}_{st}$
between $10^{-9}$ and $10^{-6}$~M$_{\odot}$~yr$^{-1}$ (assuming this
quantity is the same at all radii, i.e. the disk is in a steady state) and
for illustrative purposes we have adopted $\alpha=10^{-2}$.  In
Figures~\ref{fig2}a--b we plot the midplane temperature $T_m$ versus $r$
and $T_m$ versus the midplane pressure $P_m$ for the same parameters.  The
surface mass density of planetesimals can be derived from $\Sigma$ by
noting that in protostellar disks the gas to dust mass ration is about 100.

\noindent In these models, reprocessing of the stellar radiation by the
disk has not been included.  Figure~\ref{fig1}a indicates that the outer
parts of the disk beyond some critical radius ($\sim 0.1$ to a few AU,
depending on $\alpha$ and $\dot{M}_{st}$) are shielded from the radiation
of the central star by the inner parts (the very outer parts may not be
shielded but since they are optically thin they do not absorb the stellar
radiation anyway).  Since reprocessing of the stellar radiation by the disk
is not an important heating factor below these radii, this indicates that
this process can be self-consistently ignored.  However, it is possible
that a multiplicity of solutions exists when reprocessing is taken into
account (Chiang \& Goldreich 1997), with it being important for cases in
which the disk is flared, as may be the case with HH30 (Burrows et al.
1996), and unimportant when it is not, as may be the case with HK~Tau
(Stapelfeldt et al. 1998, Koresko 1998).  In any case, reprocessing, when
present, does not affect significantly the structure of the disk at $r \sim
1$--5~AU, where planets are supposed to form.

\noindent Below we shall consider the migration of protoplanetary cores 
from $r \sim 1$--5~AU, where they are supposed to form under conditions
where ice exists, down to the disk inner radii.  It is therefore of
interest to estimate the mass of planetesimals $M_p(r)$ contained inside
the orbit of a core located at $r$, since this can potentially be accreted
by the core during its migration.  We assume a gas to dust mass ratio of
100 and list the values of $M_p$ corresponding to $\alpha=10^{-2}$ and
$10^{-3}$ and $\dot{M} = 10^{-6}$ and $10^{-7}$~M$_{\odot}$~yr$^{-1}$
(characteristic of the early stages of disk evolution) in Table~\ref{tab1}.

\begin{table}[ht]
\caption[]{Estimate for the mass of planetesimals $M_p(r)$ contained 
within a radius $r$ for different values of $\alpha$ and 
gas accretion rate $\dot{M}$ through the disk. \\}
\begin{tabular}{cccc} 
\hline
\hline
$r$ & $\alpha$ & $\dot{M}$ & $M_p(r) = 10^{-2} \times \pi r^2 \Sigma(r)$
\\
(AU) & & (M$_{\odot}$~yr$^{-1}$) & (M$_{\oplus}$) \\ 
\hline
1   & $10^{-2}$ & $10^{-6}$ &  3.5 \\ 
... & ...       & $10^{-7}$ &  0.6 \\ 
5   & ...       & $10^{-6}$ &  22.8 \\ 
... & ...       & $10^{-7}$ &  7.6 \\ 
1   & $10^{-3}$ & $10^{-6}$ &  30.1 \\ 
... & ...       & $10^{-7}$ &  4.0 \\
5   & ...       & $10^{-6}$ &  138.5 \\ 
... & ...       & $10^{-7}$ &  41.0 \\
\hline
\end{tabular}
\label{tab1}
\end{table} 

\noindent It is also of interest, in relation to the possibility of giant 
planets being located at small radii, to estimate the mass of gas contained
within a radius of 0.1~AU.  When $\alpha= 10^{-3}$ or $10^{-2}$, this mass
is about 0.3 Jupiter mass for ${\dot M} > 10^{-7}$ or
$10^{-6}$~M$_{\odot}$~yr$^{-1}$, respectively.  For typical mass
through--put of about $10^{-2}$--$10^{-1}$~M$_{\odot},$ the lifetime of
such a state can range between $10^4$ and $10^6$~yr.  Supposing the disk to
be terminated at some small inner radius, this suggests that, if a suitable
core can migrate there, it could accrete enough gas to become a giant
planet within the disk lifetime.  We note however that the conditions for
that to happen are marginal even in the early stages of the life of the
disk when ${\dot M} > 10^{-7}-10^{-6}$~M$_{\odot}$~yr$^{-1}$.  At later
stages, when $\dot M \sim 10^{-8}$~M$_{\odot}$~yr$^{-1},$ the models
resemble conditions expected to apply to the minimum mass solar nebula with
$\Sigma \sim 200$~g~cm$^{-2}$ at 5~AU if $\alpha=10^{-2}$.  Under these
conditions, the mass of gas at $r < 0.1$~AU is between 1 and 9~M$_{\oplus}$
for $\alpha$ between $10^{-2}$ and $10^{-3}$.

\section{Migration}
\label{sec:migration}

Eight extrasolar planets among the twenty detected so far orbit at a
distance between 0.046 and 0.11~AU from their host star.  Recent studies
show that {\it in situ \/} formation of these short--period giant planets
is very unlikely (Bodenheimer 1998; Bodenheimer, Hubickyj \& Lissauer
1999).  It is therefore a possibility that these planets have formed
further away in the protoplanetary nebula and have migrated down to small
orbital distances.  It is also possible that migration and formation were
concurrent.

So far, three mechanisms have been proposed to explain the location of
planets at very short orbital distances.  One of them relies on the
gravitational interaction between two or more Jupiter mass planets, which
may lead to orbit crossing and to the ejection of one planet while the
other is left in a smaller orbit (Rasio \& Ford 1996; Weidenschilling \&
Marzari 1996).  However, this mechanism cannot account for the relatively
large number of short--period planets observed.  Another mechanism is the
so--called 'migration instability' (Murray et al. 1998; Malhotra 1972).  It
involves resonant interactions between the planet and planetesimals located
inside its orbit which lead to the ejection of a fraction of them while
simultaneously causing the planet to migrate inwards.  To move a Jupiter
mass planet from about 5~AU down to very small radii through this process,
a disk containing about 1 Jupiter mass of planetesimals, and thus about
0.1~M$_{\odot}$ of gas, inside 5~AU is required.  Such a massive disk is
unlikely and furthermore it would be only marginally gravitationally
stable.  The third mechanism, that we are going to focus on here, involves
the tidal interaction between the protoplanet and the gas in the
surrounding protoplanetary nebula (Goldreich \& Tremaine 1979, 1980; Lin \&
Papaloizou 1979, 1993 and references therein; Papaloizou \& Lin 1984; Lin
et al. 1996; Ward 1986, 1997b).  Here again the protoplanet can move
significantly only if there is at least a comparable mass of gas within a
radius comparable to that of its orbit. However this is not a problem since
this amount of gas is needed anyway in the first place to form the planet.

\subsection{Protoplanet--Disk Tidal interaction}

A protoplanet (that we shall assume is on a circular orbit) embedded in a
disk at an orbital distance $r_p$ exerts a perturbation on both sides of
its orbit.  Since the particles located at $r>r_p$ rotate slower than the
planet, they gain angular momentum when they interact with it while the
planet loses angular momentum.  As a result, these particles are pushed
further out in the disk, away from the planet which moves inwards.
Similarly, interaction between the planet and the particles located at
$r<r_p$ results in these particles being pushed further in while the planet
moves outwards.  If the interaction with the disk inner parts and that with
the disk outer parts result in exactly opposite angular momentum
exchanges, the planet does not move relative to the gas, although the
trajectory of the particles on both sides of its orbit is perturbed and a
gap may open up if the perturbation is nonlinear (Lin \& Papaloizou 1979;
Goldreich \& Tremaine 1980).  However, if the perturbation is not
symmetrical, the planet moves relative to the gas (Goldreich \& Tremaine
1980; Ward 1986, 1997b). \\

{\it Lindblad Resonances: \/} To consider the response of the disk to the
perturbation, it is convenient to expand the potential due to the
protoplanet in a Fourier series in the azimuthal angle.  The frequency of
the perturbation corresponding to the term with azimuthal mode number $m$
in a frame rotating with the fluid is $\omega_m = m(\Omega_p - \Omega)$,
where $\Omega_p$ is the planet's orbital frequency.  This perturbation can
be in resonance with the free oscillations of the disk.  As we shall see
below, the protoplanet/disk interaction is dominated by these resonances.
The locations $r_{LR}$ where $\omega_m =\pm \kappa_g$, where $\kappa_g$ if
the frequency of the free (epicyclic) oscillation of the gas, are called
Lindblad resonances.  For each value of $m$, the inner (outer) Lindblad
resonance is the radius $r_{ILR} < r_p$ ($r_{OLR} > r_p$), if it exists,
where $\omega_m=-\kappa_g$ ($\omega_m=\kappa_g$).  The location $r=r_p$
where the perturbation corotates with the fluid, i.e.  $\omega_m=0$, is
called the corotation resonance.  The perturbation propagates as density
waves outside the Lindblad resonances (which are the waves turning points),
i.e.  at $r<r_{ILR}$ and $r>r_{OLR}$, and is evanescent inside these
resonances, in the corotation region (Lin \& Shu 1964; Toomre 1981).

The protoplanet exerts a torque on the density waves, which is responsible
for the exchange of angular momentum between the disk and the planet's
orbital motion.  For a given finite $m$, the net torque exerted on both
sides of the planet's orbit is obtained by integrating the torque with
respect to radius.  Away from the Lindblad resonances, the waves have a
small wavelength so that the contribution to the integral is small.
Therefore, most of the torque is exerted in the vicinity of the Lindblad
resonances, where the perturbation has a large wavelength.  Consequently,
the total torque exerted on both sides of the planet's orbit is obtained by
summing up over $m$ the torques exerted at Lindblad resonances (Goldreich
\& Tremaine 1979). \\

{\it Torque cutoff: \/} At large $m$, the perturbing potential is more and
more localized around $r_p$ whereas the Lindblad resonances are located a
finite distance away from $r_p$.  Therefore coupling is lost, and the
contribution to the total torque becomes negligible.  

The fact that $\left| r_{LR} - r_p \right|$ stays finite as $m$ increases
seems contradictory with the definition of $r_{LR}$ we gave above.  Indeed,
the nominal Lindblad resonances defined above are such that $r_{LR}
\rightarrow r_p$ as $m$ increases.  If the effective Lindblad resonances
(i.e. the turning points of the waves, where the torque has to be
evaluated) coincided with the nominal resonances, we would have to evaluate
the torque at locations closer and closer to the planet, which would
require nonlinear calculations.  However, pressure effects cause the
location of the effective Lindblad resonances to differ from that of the
nominal Lindblad resonances when $m \wig> r/H$.  The effective inner
(outer) Lindblad resonances indeed converge towards $\sim r_p - H$ ($\sim
r_p+H$) rather than towards $r_p$ when $m \rightarrow \infty$ (Goldreich \&
Tremaine 1980; Artymowicz 1993).  Since the perturbing potential at these
locations decreases exponentially when $m$ increases (the potential is more
and more localized around $r_p$), there is a 'torque cutoff' and the
contribution to the total torque from values of $m \wig> r/H$ is
negligible.

So far we have limited the discussion to the so--called 'Lindblad torque'.
There is also a torque exerted at the position of the corotation resonance,
where the perturbation corotates with the planet.  This 'corotation torque'
should be added to the Lindblad torque (Goldreich \& Tremaine 1979). \\

{\it Type--I Migration: \/} When the perturbation exerted on the disk is
small enough to be treated using linear theory, Ward (1986; 1997b) has
shown that, because in a (even uniform) Keplerian disk the outer Lindblad
resonances are slightly closer to the planet than the inner Lindblad
resonances, the interaction with the outer parts of the disk leads to a
larger Lindblad torque than that with the inner parts.  Therefore, the net
Lindblad torque exerted by the protoplanet causes it to lose angular
momentum and to move inwards relative to the gas.  This type of migration
is referred to as 'type--I'.  In his calculations, Ward has assumed that
the corotation torque is small compared to the net Lindblad torque.  We
note that this may not be the case since, in this linear regime, the
corotation zone is not cleared up.  This point still needs to be addressed,
but here we will carry the discussion on by assuming that the corotation
torque can indeed be neglected.  Under these conditions, the drift
timescale for a planet of mass $M_{pl}$ undergoing type--I migration in a
uniform disk is (Ward 1986, 1997b):

\begin{equation}
\tau_{\rm{I}} (\rm{yr}) \sim 10^8 
\left( \frac{M_{pl}}{\rm{M}_{\oplus}} \right)^{-1} 
\left( \frac{\Sigma}{\rm{g~cm}^{-2}} \right)^{-1}
\left( \frac{r}{\rm{AU}} \right)^{-1/2}
\times 10^2 \left(\frac{H}{r} \right)^2
\label{tauI}
\end{equation}

\noindent   It is insensitive to the disk surface density profile but 
it decreases if the disk midplane temperature increases inwards.  Also
migration could reverse from inwards to outwards if the temperature
decreased inwards faster than linearly.  In a disk with $\alpha=10^{-2}$
and $\dot{M}=10^{-7}$~M$_{\odot}$~yr$^{-1}$, we have $H/r \simeq 10^{-1}$
and $\Sigma \simeq 600$~g~cm$^{-2}$ (see \S~\ref{sec:disks}), so that
$\tau_{\rm{I}} \sim 10^{5}$~yr for a 1~M$_{\oplus}$ planet at $r=1$~AU.  In
the same disk, we get $\tau_{\rm{I}} \sim 10^4$~yr for a 10~M$_{\oplus}$
planet at $r=5$~AU (where $\Sigma \simeq 300$~g~cm$^{-2}$ and $H/r \simeq
10^{-1}$).  These timescales are much shorter than the disk lifetime or
estimated planetary formation timescales.

In Ward's calculations, there is no feedback torque from the disk on the
protoplanet.  The interaction is linear, there is no gap forming, and the
evolution timescale depends only on torques acting on the unperturbed disk
structure.  However, when the mass of the protoplanet becomes large enough,
the perturbation becomes nonlinear and feedback torques cannot be
neglected.  When the protoplanet undergoes type--I migration, it pushes the
gas ahead of it in its radial drift.  This leads to an enhancement (a
depletion) of the surface mass density which leads (trails) the
protoplanet.  The feedback torque produced by this perturbed profile of the
surface mass density opposes the motion of the protoplanet, and stops it
altogether when the planet is massive enough to perturb significantly the
distribution of mass, i.e. when the perturbation becomes strongly nonlinear
(Ward 1997b).  Thus type--I migration occurs only for small mass planets.
Because the viscosity tends to smooth out perturbations of the density
profile, it decreases the feedback torque.  Therefore, the larger the
viscosity, the larger the mass of the planet for which type--I migration is
stopped (see Figure~14 from Ward 1997b). \\

{\it Type--II Migration: \/} Once the perturbation exerted by the
protoplanet becomes nonlinear, it affects the structure of the disk around
its orbit.  If the density waves excited by the protoplanet are dissipated
locally, the angular momentum they transport is deposited into the disk in
the vicinity of the protoplanet, and a gap may be cleared out (Goldreich \&
Tremaine 1980; Lin \& Papaloizou 1979, 1993 and references therein).  The
strongest Lindblad torques are exerted at a distance $\sim H$ from the
protoplanet (at the location of the Lindblad resonances corresponding to $m
\sim r/H$).  Therefore, a gap of width $\sim H$ will open up if the waves
launched at these locations dissipate before they can propagate
significantly.  If dissipation is due to nonlinearity of the waves, this
requires (Lin \& Papaloizou 1993; Korycansky \& Papaloizou 1996; Ward
1997b):

\begin{equation}
\frac{M_{pl}}{M_{\ast}} \wig> 3 \left( \frac{H}{r} \right)^3 .
\label{crit1}
\end{equation}

\noindent In order for the gap to be maintained, the tidal torque at its
edges must exceed the intrinsic viscous torque, provided the tidally
excited waves are completely damped.  This requires (Lin \& Papaloizou
1979):

\begin{equation}
\frac{M_{pl}}{M_{\ast}} \wig> 40 \alpha \left( \frac{H}{r} \right)^2 .
\label{crit2}
\end{equation}

\noindent Once these two conditions are met, transfer of mass through the gap 
is reduced, or even switched off altogether (Kley 1999; Bryden et al.
1999).  However, transfer of angular momentum from the disk inner parts to
the protoplanet, and from the protoplanet to the disk outer parts is still
taking place.  Therefore, the protoplanet is locked into the angular
momentum transport process of the disk.  Unless the protoplanet is located
in the disk outer parts, which may be diffusing outwards (Lynden--Bell \&
Pringle 1974), it is going to migrate inwards at a rate controlled by the
disk viscous timescale:

\begin{equation}
\tau_{\rm{II}} (\rm{yr}) = \frac{1}{ 3\alpha} \left( \frac{r}{H} \right)^2
\Omega^{-1} = 0.05 \frac{1}{\alpha} \left( \frac{r}{H} \right)^2 \left(
  \frac{r}{\rm{AU}} \right)^{3/2} .
\end{equation}
 
\noindent This type of migration is referred to as 'type--II'.  If we 
adopt $H/r=10^{-1}$ and consider $\alpha$ in the range
$10^{-2}$--$10^{-3}$, we get $\tau_{\rm{II}} \sim 5\times 10^2$--$5\times
10^3$ or $6\times 10^3$--$6\times 10^4$~yr at $r=1$ or 5~AU, respectively.
These timescales again are much shorter than the disk lifetime or estimated
planetary formation timescales.  The expression of $\tau_{\rm{II}}$ is
independent of the protoplanet's mass.  We note however that the above
discussion implicitly assumes that the Roche lobe of the planet is not
significantly larger than $H$, i.e. $M_p/M_{\ast}$ is larger than but still
comparable to $3 (H/r)^3$.  If it were not the case, the edges of the gap
would be set by the protoplanet's Roche lobe or the location of the 2:1
resonance rather than by the location of the Lindblad resonances
corresponding to $m \sim r/H$ (Lin \& Papaloizou 1979).  Also the surface
mass density of the disk does not appear in the above timescale, but it is
implicitly assumed here that the mass of gas within the characteristic
orbital radius of the planet is at least comparable to the mass of the
planet itself.

In the context of a nonlinear perturbation, what would happen if the
Lindblad torque on one side of the planet's orbit were larger than that on
the other side, as in the case leading to type--I migration ?  Once a gap
has formed, an unbalanced Lindblad torque would push the planet towards the
edge of the gap at which the torque is smaller, strengthening the torque
on that side until balance is attained.  The only equilibrium solution is
indeed one where ultimately the torques exerted on either side are equal
and opposite.  So only the position of the planet inside the gap would be
affected by an initially finite net torque.

As pointed out above, a clean gap can be cleared out only if the
perturbation is strongly nonlinear.  If the perturbation is not strong
enough for the corotation region to be completely emptied out (regime that
we shall call 'transitional'), we have only partial gap clearing and the
evolution is controlled both by the disk viscosity and the action of the
torques on the partially perturbed disk. \\

{\it Summary: \/} In table~\ref{tab2}, we summarize our present knowledge
of the different stages of migration the protoplanet goes through as its
mass $M_{pl}$ increases.  We indicate whether the interaction is linear or
not, whether there is a feedback from the disk and a subsequent gap
formation, which type of migration the protoplanet is undergoing, and
whether the drift timescale is controlled by the torques acting on the disk
and/or the disk viscosity.  It is worth pointing out again that many
simplifying assumptions are made in order to do these calculations, and
that in particular some issues related to type--I migration still need to
be addressed.

\begin{table}[ht]
\caption[]{Different stages of migration for a protoplanet of 
mass $M_{pl}$. \\}
\begin{tabular}{ccccccccccccccccccccccccccc} 
\hline
\hline
$M_{pl}$ & Interaction & Gap, disk & Migration & Evolution timescale  \\
         &             & feedback  &           & controlled by  \\
\hline
Small & Linear & No & {\bf Type--I}, {\it relative}  & Torques \\
      &        &    &  to the disk                   &         \\
Intermediate & Nonlinear & Partial & {\bf Type I/II} & Torques and \\
             &           &         &                 &  viscosity  \\ 
Large & Strongly  & Strong & {\bf Type--II}, {\it with} & Viscosity \\
      & nonlinear &        & the disk                   &           \\

\hline
\end{tabular}
\label{tab2}
\end{table} 

\subsection{How to stop the migration ?}

The previous section indicates that migration is relatively fast.  In the
context of the disk models presented above, a core which builds up at a
location between 1 and 5~AU will migrate inwards once it reaches a mass of
about 0.1~M$_{\oplus}$ on a timescale shorter than the time it takes to
form a giant planet, according to current theories (Pollack et al.  1994).
Therefore, unless there is a mechanism to stop the migration of this core,
it will never become a giant planet but will plunge onto the star.  Type--I
migration could be avoided if the interaction could become nonlinear and a
gap open up around the core.  This would require small values of $H/r$ and
$\alpha$ (see eq.~[\ref{crit1}] and~[\ref{crit2}]) and then type--II
migration would be very slow.  If a gap opens up around a core before it
has accreted a gaseous envelope, further accretion may take place (e.g.,
Kley 1999; Bryden et al. 1999) so that it is conceivable it could become a
giant planet.  However, observations show that in some cases migration has
to take place to push the protoplanet down to intermediate radii ($\sim
0.5$~AU), and has to stop there.  If the viscosity were controlling the
migration rate, unless there is a fortuitous disk dispersal, this would
require $\alpha$ to decrease as $r$ decreases, in contrast to what is
proposed in current disk models (see Balbus \& Hawley in this volume).  In
the context of the disk models presented in \S~\ref{sec:disks}, only a low
surface mass density of gas would give long enough migration timescales to
enable planet formation to take place before protoplanets with masses $>
0.1M_{\oplus}$ fall onto their parent star.

In this context, Ward (1997a) has suggested that several small mass ($\wig<
0.1$~M$_{\oplus}$) cores could be built--up and maintained in isolation
until the disk is partially depleted, at which point they would assemble
into a larger core.  However, it seems that such cores would interact with
each other on a rather short timescale (Chambers, Wetherill \& Boss 1996).
Also, even if the core could assemble only in the late stages of the disk
evolution, there may no longer be enough gas left in the disk at this point
to form a giant planet, let alone a system of three giant planets, as
observed around Ups And (Butler et al. 1999).

Therefore either giant planet formation occurs faster than currently
thought, on a timescale shorter than $\tau_{\rm{I}}$, or type--I migration
can be halted somehow.  We note that migration would cease if the disk were
terminated by a magnetospheric cavity and the core were sufficiently far
inside it (Lin et al. 1996).  However, gas accretion is likely to be very
much reduced in this case.  The core migration should be halted before the
disk is terminated for a giant planet to be formed.  As noted above, Ward
(1986, 1997b) finds the direction of type--I migration could reverse from
inwards to outwards if the disk midplane temperature decreased inwards fast
enough.  Such a condition would not be expected in the disk models
considered here.  However, if the inner disk is terminated through
interaction with a stellar magnetic field, physical conditions may start to
differ in the interaction zone where magnetic field lines penetrate the
disk.  Additional energy and angular momentum transport mechanisms due to a
wind for example may start to become important.  As a result, an inward
midplane temperature decrease might be produced.  It may then be possible
that migration could be halted such that the protoplanet retains contact
with disk gas.

In addition, work in progress (the results of which will be published
elsewhere) shows that type--I migration is affected by the presence of a
magnetic field in the disk.  As described above, the torque exerted by the
protoplanet on the disk depends mainly on the location of the radii where
the perturbation is in resonance with the free oscillations of the disk.
When a magnetic field is present, the perturbation can resonate not only
with the epicyclic motions of the fluid but also with the Alfven and
magnetoacoustic waves which propagate through the fluid.  Therefore there
are additional locations in the disk where a torque can be exerted.  This
additional contribution to the torque has to be considered.  Even if it
were not important, the results derived by Ward (1986, 1997b) and described
above would be affected.  Whether the outer Lindblad resonances are closer
or not to the protoplanet than the inner Lindblad resonances indeed depends
on the gradient of the Lorentz force in the vicinity of the protoplanet.

Also, preliminary estimates of the effect of a finite eccentricity $e$ of
the planet's orbit on type--I migration indicate that it may be reversed
for reasonable disk models once $e \wig> H/r$.  Details will be published
elsewhere.

\section{Formation of a Giant Planet  with a Migrating Core}

Giant planets are believed to form out of protostellar disks by either
gravitational instability (Kuiper 1951; Cameron 1978; Boss 1998) or by a
process of growth through planetesimal accumulation followed by gas
accretion (Safronov 1969; Wetherill~\& Stewart 1989; Perri~\& Cameron 1974;
Mizuno 1980; Bodenheimer~\& Pollack 1986).  The first mechanism is expected
to produce preferentially massive objects in the outer parts of the disk,
if anything.  Here we study planetary formation within the context of the
second mechanism, which is commonly accepted as the most likely process by
which planets form in at least the inner ten astronomical units of
protostellar disks.  We note however that important issues related to this
model still remain to be resolved (see Lissauer 1993 for a review).

The build--up of the atmosphere of giant planets has first been considered
in the context of the so--called 'core instability' model by Perri~\&
Cameron (1974) and Mizuno (1980; see also Stevenson 1982 and Wuchterl
1995).  In this model, the solid core grows in mass along with the
atmosphere in quasi--static and thermal equilibrium until the core reaches
the so--called 'critical core mass' $M_{crit}$ above which no equilibrium
solution can be found for the atmosphere.  As long as the core mass
$M_{core}$ is smaller than $M_{crit}$, the energy radiated from the
envelope into the surrounding nebula is compensated for by the
gravitational energy which the planetesimals entering the atmosphere
release when they collide with the surface of the core.  During this phase
of the evolution, both the core and the atmosphere grow in mass relatively
slowly.  By the time $M_{core}$ reaches $M_{crit}$, the atmosphere has
grown massive enough so that its energy losses can no longer be compensated
for by the accretion of planetesimals alone.  At that point, the envelope
has to contract gravitationally to supply more energy.  This is a runaway
process, leading to the very rapid accretion of gas onto the protoplanet
and to the formation of giant planets such as Jupiter.  Further
time--dependent numerical calculations of protoplanetary evolution by
Bodenheimer~\& Pollack (1986) support this model, although they show that
the core mass beyond which runaway gas accretion occurs, which is referred
to as the 'crossover mass' (because it is reached when the mass of the
atmosphere is comparable to that of the core), is slightly larger than
$M_{crit}$.  The similarity between the critical and crossover masses is
due to the fact that, although there is some liberation of gravitational
energy as the atmosphere grows in mass together with the core, the effect
is small as long as the atmospheric mass is small compared to that of the
core.  Consequently, the hydrostatic and thermal equilibrium approximation
for the atmosphere is a good one for core masses smaller than $M_{crit}$.
Therefore we use this approximation here and investigate how $M_{crit}$
varies with location and physical conditions in the protoplanetary disk.

\subsection{Critical core mass as a function of location and disk 
parameters}

To calculate the structure of an atmosphere in hydrostatic and thermal
equilibrium around a core of mass $M_{core}$, we need to solve the
equations of hydrostatic equilibrium, mass conservation and radiative
transport (Mizuno 1980).  Both radiative and convective transport are taken
into account here, and it is assumed that the only energy source comes from
the accretion of planetesimals onto the core.  The atmosphere is confined
between the radius of the core and that of the protoplanet's Roche lobe.
Details of the calculations presented here are given in Papaloizou \&
Terquem (1999), who used the equation of state given by Chabrier et al.
(1992) and the same opacity law as in \S~\ref{sec:disks}.

\begin{figure}
\centerline{\epsfig{file=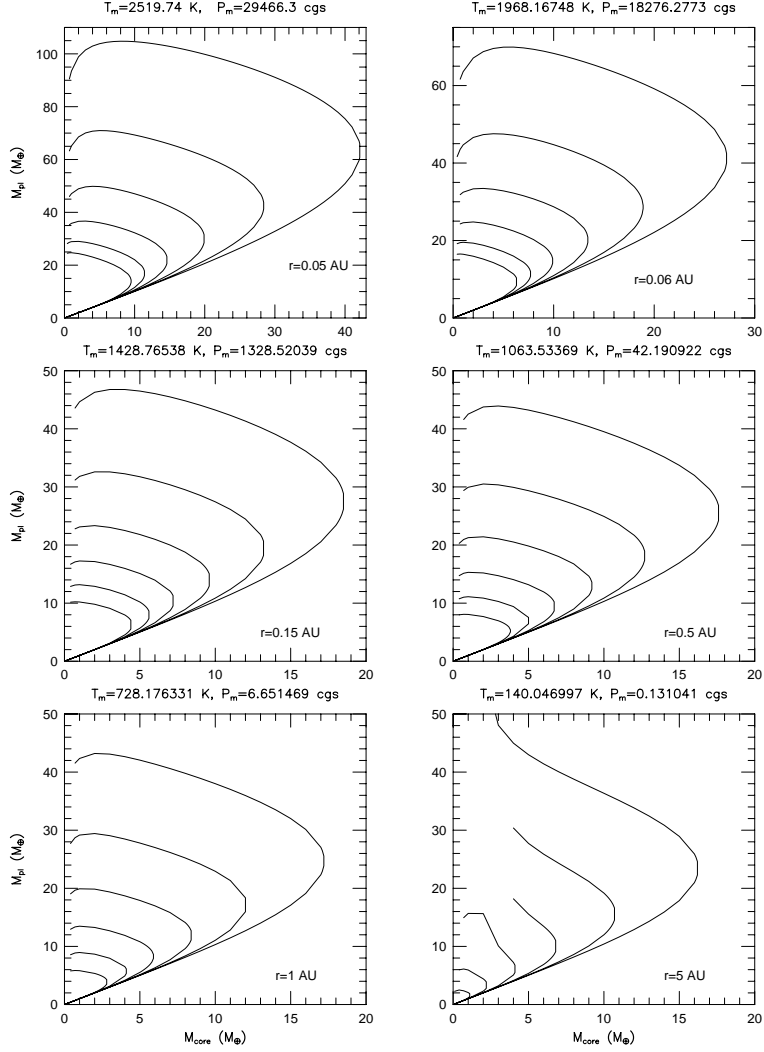,width=13.cm}}
\caption[]{Plots of total mass, $M_{pl}/{\rm{M}}_{\oplus}$,
  vs. core mass, $M_{core}/{\rm{M}}_{\oplus}$, at different locations $r$
  in a steady state disk model with $\alpha=10^{-2}$ and gas accretion rate
  ${\dot M} = 10^{-7}$~M$_{\odot}$~yr$^{-1}$.  From left to right and top
  to bottom, the frames correspond to $r=0.05$, 0.06, 0.15, 0.5, 1 and
  5~AU, respectively.  The midplane temperature and pressure at these
  locations are indicated above each frame.  Each frame contains six curves
  which, moving from left to right, correspond to core luminosities derived
  from planetesimal accretion rates of $\dot{M}_{core}=10^{-11}$,
  $10^{-10}$, $10^{-9}$, $10^{-8}$, $10^{-7}$ and
  $10^{-6}$~M$_{\oplus}$~yr$^{-1}$, respectively.  $M_{crit}$ is attained
  when the curves first begin to loop backwards when moving from left to
  right.}
\label{fig3}
\end{figure}

In Figure~\ref{fig3}, we plot curves of total protoplanet mass $M_{pl}$
against $M_{core}$ at different radii in a disk with $\alpha= 10^{-2}$ and
${\dot M}=10^{-7}$~M$_{\odot}$~yr$^{-1}$ (using the models described in
\S~\ref{sec:disks}).  In each frame, the different curves correspond to
planetesimal accretion rates $\dot{M}_{core}$ in the range
$10^{-11}$--$10^{-6}$~M$_{\oplus}$~yr$^{-1}$.  The critical core mass
$M_{crit}$, above which the equations have no solution, is attained at the
point where the curves start to loop backwards.  For masses below
$M_{crit}$, there are (at least) two solutions, corresponding to a
low--mass and a high--mass envelope, respectively.  When the core first
begins to gravitationally bind some gas, the protoplanet is on the left on
the lower branch of these curves.  Assuming $\dot{M}_{core}$ to be
constant, as the core and the atmosphere grow in mass, the protoplanet
moves along the lower branch up to the right, until the core reaches
$M_{crit}$.  At that point the atmosphere begins to undergo very rapid
contraction.  Since the atmosphere in complete equilibrium is supported by
the energy released by the planetesimals accreted onto the protoplanet,
$M_{crit}$ decreases as $\dot{M}_{core}$ is reduced.

For $\alpha=10^{-2}$ and $\dot{M}=10^{-7}$~M$_{\odot}$~yr$^{-1}$,
$M_{crit}$ at 5~AU varies between 16.2 and 1~M$_{\oplus}$ as
$\dot{M}_{core}$ varies between the largest and smallest value.  The former
result is in good agreement with that of Bodenheimer~\& Pollack (1986).
Note that there is a tendency for $M_{crit}$ to increase as $r$ decreases,
the effect being most marked at small radii.  At 0.05AU, $M_{crit}$ varies
from 42 to 9~M$_{\oplus}$ as $\dot{M}_{core}$ varies between the largest
and smallest value.  Similar calculations for different disk models (see
Papaloizou \& Terquem 1999) indicate a relatively weak dependence of
$M_{crit}$ on disk conditions except when rather high midplane temperatures
$T_m > 1,000$~K are attained, as in the inner regions.  These results are
consistent with the fact that $M_{crit}$ depends on the boundary conditions
only when a significant part of the envelope is convective (Wuchterl 1993),
being larger for larger convective envelopes (Perri~\& Cameron 1974).  We
note that it is unlikely there are planetesimals at radii $\wig< 0.1$~AU.
Therefore, although critical core masses for the same planetesimal
accretion rates may be higher there, a lack of planetesimals may result in
a fall in the core luminosity, making $M_{crit}$ relatively small at these
radii.

\subsection{Protoplanet Migration and Planetesimal Accretion}

As noted in \S~\ref{sec:migration}, cores of several earth masses form at
about 5~AU and migrate inwards to small radii before they can become a
giant planet.  In doing so, they continue to grow.  To know whether they
can reach the critical mass on their way, and thus begin to accrete
significant amount of gas, we have to evaluate $\dot{M}_{core}$, i.e. the
mass of planetesimals they accrete as they migrate in.

Papaloizou \& Terquem (1999) performed analytical estimates and numerical
calculations to evaluate the fraction of planetesimals accreted by a core
of mass $\sim 1$--10~M$_{\oplus}$ migrating inward from $r \sim 1$~AU on a
timescale $\sim 10^3$--$10^4$~yr.  They found that the migrating
protoplanet accretes about 25\% of the planetesimals inside its orbit, as
if it were in a homogeneous medium without a gap forming in the
planetesimal distribution.  Even more planetesimals are accreted when the
migration is slower.  Given that the disk models typically contain at least
about 8~M$_{\oplus}$ interior to 5~AU (see \S~\ref{sec:disks}), this means
that the accretion of planetesimals is high enough to maintain the core
luminosity such that attainment of $M_{crit}$ (and therefore gas accretion)
does not occur as long as planetesimals are present, i.e. above $\sim
0.05$--0.1~AU.  Runaway gas accretion onto a small core can then occur at
these radii.  However, if the core is too small, the gas accretion phase
may be longer than the disk lifetime.  Even for core masses in the range
15--20~M$_{\oplus}$, the build--up of a massive atmosphere may take a time
$\sim 10^6$~yr (Bodenheimer et al.  1998).  The reason for this is that
once the core starts to accrete a significant atmosphere, energy production
occurs through its gravitational contraction.  The luminosity produced then
slows down the evolution.  An estimate of the evolutionary
(Kelvin--Helmholtz) timescale $t_{KH}$ at this stage has been calculated by
Papaloizou \& Terquem (1999).  Typically, they found $t_{KH} \sim
10^6-10^7$~yr for core masses in the range 10--20~M$_{\oplus}$ for radii
larger than 0.075~AU.  The core masses required to get such a
characteristic timescale increase rapidly interior to 0.06~AU.  However, we
note that they decrease as the mass transfer rate through the disk does.
The fact that fairly large core masses are required to give evolutionary
timescales comparable or less than the expected disk lifetime means that
mergers of additional incoming cores may be required in order to produce a
core of sufficient mass that real runaway gas accretion may begin.

Table~\ref{tab1} indicates that, in a disk with $\alpha=10^{-3}$ and ${\dot
  M} = 10^{-6}$ or $10^{-7}$~M$_{\odot}$~yr$^{-1}$, 40~M$_{\oplus}$ of
planetesimals are contained within 1 or 5~AU.  Therefore, the timescale for
building--up a core with a mass between 20 and 40~M$_{\oplus}$ at small
radii is typically the timescale it takes for planetesimals to migrate from
1 or 5~AU down to these small radii.  According to Ward (1997b; and see
\S~\ref{sec:migration}), the migration timescale of cores of $\sim
0.1$~M$_{\oplus}$ located at 1 or 5~AU is at most $10^6$~yr in such a disk,
and it decreases with increasing core mass.  If the disk has
$\alpha=10^{-2}$, 40~M$_{\oplus}$ of planetesimals are contained within 5
or 11~AU.  In this case again, the migration timescale of cores of $\sim
0.1$~M$_{\oplus}$ at 5 or 11~AU is about $10^6$~yr.  Therefore, if
planetesimals can be assembled into cores of at least a few tenths of an
earth mass at these radii on a reasonably short timescale, a massive core
could be obtained at small radii (if at least three cores are present, they
cannot stay in isolation but interact with each other, as shown by
Chambers, Wetherill \& Boss 1996).  A massive core can be built--up through
the merger of additional incoming cores either after having stopped at
small radii or on its way down to small radii (where it would still be
expected to be stopped).  The former process resembles that discussed by
Ward (1997a).  The latter scenario would occur if more massive cores, which
migrate faster, overtake less massive cores on their way down.  In any
case, several cores would necessarily interact with each other at some
point so that core isolation, which would increase planet formation very
much if it occurred (Pollack et al.  1994), would be avoided.

Supposing that a protoplanetary core massive enough can be built--up on its
way down to small radii and that it continues to rapidly move inward until
it gets interior to the disk inner boundary, it can only accrete the gas
which is in its vicinity, i.e. typically the amount of gas contained within
$\sim 0.05$--0.1~AU.  Since the core is expected to reach these radii early
in the life of the protoplanetary disk, there may still be an adequate
amount of gas there (see \S~\ref{sec:disks}) for it to build--up a large
envelope and become a giant planet.  However the conditions for that to
happen are rather marginal.  If the protoplanetary core is stopped at some
small radius before the disk is terminated (see \S~\ref{sec:migration}), it
might be able to retain contact with disk gas.  In that case it might be
able to accrete enough gas supplied from the outer disk by viscous
evolution to build--up a massive atmosphere.

The scenario discussed in this section might be able to produce short
period planets in the early stages of the disk evolution.  It would more
likely result in a single planet at $\sim 0.05$--0.1~AU on a timescale
significantly shorter than for {\it in situ \/} formation if the core were
in isolation in this latter process.

\section{Summary}

In this paper we have described briefly standard $\alpha$--disk models and
presented some numerical calculations for $\alpha=10^{-2}$ and gas
accretion rates through the disk varying from $10^{-9}$ to
$10^{-6}$~M$_{\odot}$~yr$^{-1}$. Then we have discussed planet migration
and formation in the context of these models.

We have reviewed the current theories of protoplanet migration, focusing
on the gaseous disk--protoplanet tidal interaction.  According to these
theories, protoplanets migrate from the location where they begin to form,
which is supposed to be $\sim 1$--5~AU, down to the disk inner parts, on a
timescale significantly shorter than the disk lifetime  or even the
planetary formation timescale.  Clearly the theory has to be developed to
explain how at least some of the planets halt their migration before they
plunge onto their parent star.  However, these results suggest that
planets may not form {\it in situ \/} but more likely form at the same time
as they migrate.

Accordingly, we have considered both the growth of the core and the
built--up of the envelope of a giant planet in the course of its migration.
Our calculations indicate that accretion of planetesimals during the
migration is likely to be high enough to maintain the core luminosity such
that attainment of the critical core mass (and therefore significant gas
accretion) does not occur as long as planetesimals are present, i.e. above
$\sim 0.05$--0.1~AU.  Although runaway gas accretion can then begin onto
small mass cores at these small radii, the timescale for building--up a
massive envelope becomes longer than the disk lifetime if the core is too
small.  However, cores massive enough can be built--up through mergers of
additional incoming cores on a timescale shorter than for {\it in situ \/}
formation if the core were in isolation in this latter process.  The above
considerations can lead to the preferential formation of short--period
planets on a relatively short timescale in the early stages of the disk
evolution.

\begin{acknowledgements} 

We thank S. Balbus for useful comments on an early draft of this paper.
CT is supported by the Center for Star Formation Studies at NASA/Ames
Research Center and the University of California at Berkeley and Santa
Cruz, and in part by NSF grant AST--9618548.

\end{acknowledgements}

\end{article} 
\end{document}